\documentclass[11pt]{article}
\usepackage[numbers]{natbib}
\usepackage{graphicx,epsfig}
\usepackage{latexsym}
\usepackage{amsfonts,amssymb}
\usepackage{amsmath}
\usepackage{youngtab}
\usepackage{natbib}
\usepackage{bm}
\def \TL{{\mathrm{L}}}

\def \pd{\partial}

\def \tl#1{\overset{\kern 1pt\circ}{#1}}
\def \TL#1{\overset{\kern -3pt \circ}{#1}}
\def \TLL#1{\overset{\kern -7pt \circ}{#1}}
\def \relstack#1#2{\mathrel{\mathop{#2}\limits_{#1}}}

\def\negenspace{\kern-1.1em}
\def\quer{\negenspace\nearrow}
\def\negenspaceexp{\kern-0.5em}

\newcommand\cF{\overset{\kern 1pt \circ}{F}\vphantom{F}}

\begin{document}
%%%%%%%%%%%%%%%%%%%%%%%%%%%%%%%%%%%%%%%%%%%%%%%%%%%%%%%%%%%%%%%%%%%%%

\title{{\bf
Irreducible decomposition of strain gradient tensor in 
isotropic strain gradient elasticity
}}
\author{
 Markus Lazar~$^\text{}$\footnote{
{\it E-mail address:} lazar@fkp.tu-darmstadt.de (M.~Lazar).}
\\ \\
${}^\text{}$
        Heisenberg Research Group,\\
        Department of Physics,\\
        Darmstadt University of Technology,\\
        Hochschulstr. 6,\\
        D-64289 Darmstadt, Germany\\
}

\date{\today}
\maketitle

%%%%%%%%%%%%%%%%%%%%%%%%%%%%%%%%%%%%%%%%%%%%%%%%%%%%%%%%%%%%%%%%%%%%%%%%%%%%%
\begin{abstract}
In isotropic strain gradient elasticity, we decompose the strain gradient tensor into its irreducible pieces under the
$n$-dimensional orthogonal group $O(n)$.   
Using the Young tableau method for traceless tensors, four irreducible pieces ($n>2$), which are canonical, are obtained.     
In three dimensions, the strain gradient tensor can be decomposed into four irreducible pieces with 7+5+3+3
independent components whereas in two dimensions, the strain gradient tensor can be decomposed into three irreducible pieces with 2+2+2 independent components. 
The knowledge of these irreducible pieces is extremely  useful when setting up
constitutive relations and strain energy. 
\\

\\
\end{abstract}

%\end{frontmatter}

\section{Introduction}
Irreducible tensors are of fundamental interest in the representation theory of groups and are important in applied mathematics and theoretical physics (e.g., \citep{Weyl1,Weyl,Hamer,BR}). 
Irreducible tensors of rank three and four are used in differential geometry and metric affine gauge theory 
(e.g., \citep{Hehl95}). 
In high energy physics, tensors of arbitrary rank being
irreducible with respect to the Lorentz group SO(1,3) are necessary (e.g., \citep{GLR99,GL00}). 	
In generalized elasticity, \citet{Toupin62} discussed that general tensors of rank three
have four irreducible symmetry parts. 
\citet{Toupin62} used four symmetry operators, which correspond to the four
relevant Young symmetries (symmetries of Young tableaux, see below).  
The four irreducible symmetry parts discussed by~\citet{Toupin62}
are irreducible with respect to the $n$-dimensional general linear group $GL(n)$.
For couple-stress elasticity, only two irreducible symmetry parts survive
since the couple-stress tensor is antisymmetric in two indices.
Currently, there is a renewed interest in irreducible tensors 
in elasticity theory (e.g., \citep{IH}) and in generalized elasticity theory,
especially in gradient elasticity theory (e.g., \citep{Auffray13,Auffray15,GK}).

Mindlin's theory of strain gradient 
elasticity~\citep{Mindlin64} is a well-suited framework 
 to model the behavior of elastic materials up to the nano-scale. 
 Using ab initio calculations, 
\citet{Shodja} found that the characteristic length scale parameters of Mindlin's
gradient elasticity theory are in the order of $\sim 10^{-10}$~m 
for several fcc and bcc materials. Therefore, as a generalization of classical elasticity, gradient elasticity is relevant for nano-mechanical phenomena at such length scales.
However, the most general version of Mindlin's strain gradient elasticity  has
found limited application so far, because of both 
its complexity and the presence of a large number of  new
material parameters.
The use of irreducible tensors in Mindlin's strain gradient elasticity theory may deliver a better understanding
of the general structure of such a theory.  

The aim of this work is to give a canonical and unique decomposition of
the strain gradient tensor, which is a tensor of rank three, into its
irreducible pieces with respect to the orthogonal group $O(n)$ using the
method of Young tableaux. Also the connection 
of the irreducible form of strain gradient elasticity to Mindlin's strain 
gradient elasticity theory \citep{Mindlin64} and to 
strain gradient elasticity theory of Helmholtz type \citep{LM05,Lazar14} is given.

\section{Prolog}

In {\it strain gradient elasticity} (or {\it gradient elasticity of form II}), 
the {\it two state quantities} are the 
{\it elastic strain tensor}, which is a symmetric tensor of rank two, 
\begin{align}
\varepsilon_{ij}=\frac{1}{2}\, \big(u_{i,j}+u_{j,i}\big)\,, 
\qquad
\varepsilon_{ij}=\varepsilon_{ji}
\end{align}
and the {\it elastic strain gradient tensor}, which is a tensor of rank three, 
\begin{align}
\eta_{ijk}:=\varepsilon_{ij,k}\,, \qquad \eta_{ijk}=\eta_{jik}
\end{align}
is symmetric in the first two indices\footnote{We use the notation of \citet{Schouten}. 
Symmetrization over two indices is denoted by parentheses, $A_{(ij)}:=(A_{ij}+A_{ji})/2!$,
antisymmetrization by brackets, $B_{[ij]}:=(B_{ij}-B_{ji})/2!$. The analogous is valid for more indices, e.g.,
$C_{(ijk)}:=(C_{ijk}+C_{jki}+C_{kij}+C_{ikj}+C_{jik}+C_{kji})/3!$.}:
$\eta_{ijk}\equiv\eta_{(ij)k}$. 
Here $u_i$ is the {\it displacement vector} and a comma denotes 
the differentiation with respect to the coordinates. 

The elastic strain tensor can be decomposed into two irreducible pieces according to
\begin{align}
\varepsilon_{ij}&= {}^{(1)}\varepsilon_{ij}+{}^{(2)}\varepsilon_{ij}
%%\frac{1}{n}\, \varepsilon_{\cdot\cdot}
\end{align}
with the number of independent components
\begin{align}
\quad
\frac{n(n+1)}{2}=\frac{(n+2)(n-1)}{2}+1\,.
\end{align}
The irreducible tensor piece ${}^{(1)}\varepsilon_{ij}$ is defined as symmetric and
traceless tensor of rank two, which is the {\it shear tensor} 
%%%{\varepsilon\quer}_{ij}$ is defined as symmetric and traceless tensor of rank two
\begin{align}
 {}^{(1)}\varepsilon_{ij}=
{\varepsilon\quer}_{\!\! ij}:=\varepsilon_{ij}-\frac{1}{n}\,\delta_{ij}\, \varepsilon_{\cdot\cdot}\,, \qquad
{\varepsilon\quer}_{\!\!\cdot\cdot}=0\,.
\end{align}
In tensor analysis and continuum mechanics, this piece is usually called {\it
  deviator}  of $\varepsilon_{ij}$ (e.g.,~\citep{Bertram,Chadwick}).
The second  irreducible tensor piece ${}^{(2)}\varepsilon_{ij}$ 
is the {\it spherical part} of $\varepsilon_{ij}$
\begin{align}
{}^{(2)}\varepsilon_{ij}=\frac{1}{n}\, \delta_{ij}\, \varepsilon_{\cdot\cdot}
\end{align}
with the {\it dilatation} $\varepsilon_{\cdot\cdot}$
\begin{align}
\varepsilon_{\cdot\cdot}:=\varepsilon_{ii}=\delta_{ij}\,\varepsilon_{ij}
\end{align}
%%is the {\it spheric part} of $\varepsilon_{ij}$
and $\delta_{ii}=n$.

\section{Irreducible decomposition of the strain gradient tensor $\eta_{ijk}$}
\subsection{Basic remarks}

Let $O(n)$ be the orthogonal group in $n$-dimensions.
 {\it $O(n)$-irreducible tensors are traceless tensors having defined symmetry classes} associated with 
so-called {\it Young tableaux}  (see Appendix~\ref{young} 
and \citep{Hamer,Weyl,Boerner}).
Starting from the subspace of traceless tensors and applying the Young symmetry to obtain irreducible tensors of
a given symmetry type.  
However, not all Young tableaux are admissible.
The {\it traceless tensors corresponding to Young tableaux where the sum of the lengths of the first two columns exceeds
n must be identically zero} \citep{Hamer}. 
Thus, only the Young tableaux are permissible for which the sum of the lengths of the first two
columns is: $\mu_1+\mu_2\le n$.
Two tableaux $T$ and $T'$ (and the corresponding tensor representations) 
for which the first columns are related by $\mu_1'=n-\mu_1$,
where $\mu_1\le n/2$, are called {\it associated}.
If $\mu_1=\mu_1'=n/2$ ($n$ is even), 
then $T$ and $T'$ are called {\it self-associated}. 
If we consider the special (or proper) orthogonal group $SO(n)\subset O(n)$, 
for which $\bm a\in SO(n)$: $\text{det} \,\bm a =+1$, 
the representations corresponding to the associated Young pattern $T$ and $T'$ become
{\it equivalent}, 
whereas representations corresponding to self-associated pattern 
decompose into two {\it nonequivalent} irreducible representations.
As an  example for $n=3$,
the pattern $T=(1)$ and $T'=(1,1)$ are associated:
\begin{eqnarray}\label{tab1}
 \Yvcentermath1\young({\ })\qquad &&
 \Yvcentermath1\young({\ },{\ }) \nonumber\\
T=(1) && T'=(1,1)
\end{eqnarray}
The pattern $T=(1)$ describes vectors, e.g. $V_i$, 
while $T'=(1,1)$ describes antisymmetric tensors of rank two, e.g. $T_{[jk]}$. 
The vector and axial vector representations $V_i$ and $A_i=\frac{1}{2}\, \epsilon_{ijk}T_{[jk]}$,
respectively, are associated  ones ($\epsilon_{ijk}$ is the three-dimensional
Levi-Civit\`a tensor). 
Under $SO(3)$ both types transform in the same way and $T$ and $T'$ are equivalent. 
Under $O(3)$, the tensors for $T=(1)$ change sign (polar vector), whereas 
$T'=(1,1)$ do not (axial vector).

\subsection{Young symmetries of a general tensor of rank three}
\label{YS-T3}

For a {\it general tensor of rank three},  $F_{ijk}$,
there are {\it four non-vanishing standard Young tableaux} if $n>2$, namely 
\begin{eqnarray}\label{tabT3}
 \Yvcentermath1\young(i)\otimes\young(j)\otimes\young(k)
 &=&\Yvcentermath1 \alpha\,\young({i}{j}{k})
 \Yvcentermath1 + \beta\Bigg(\young({i}{j},{k}) 
+ \young({i}{k},{j}) \Bigg)
 + \gamma\,  \young({i},{j},{k}) \,
%\end{align}
\end{eqnarray}
where the coefficients $\alpha$, $\beta$, $\gamma$ are determined by the 
formula~(\ref{f}) and they are (see also~\citep{Hamer})
\begin{align}
\label{factors}
\alpha=\frac{f_{[3]}}{3!}=\frac{1}{3!}\,,\qquad
\beta=\frac{f_{[2,1]}}{3!}=\frac{2}{3!}\,,\qquad
\gamma=\frac{f_{[1,1,1]}}{3!}=\frac{1}{3!}\,.
\end{align}
Thus, for a general tensor of rank three $F_{ijk}$, the Young tableaux
in~(\ref{tabT3}) with (\ref{factors}) give the following tensor decomposition 
\begin{align}
\label{deco-YT-T3}
F_{ijk}&={}_{\text{S}}F_{ijk}+{}_{\bar{\text{P}}}F_{ijk}+{}_{\text{P}}F_{ijk}+{}_{\text{A}}F_{ijk}
\end{align}
where 
%%%% $\alpha=1/12$.
\begin{align}
\label{TS}
{}_{\text{S}}F_{ijk}&=\frac{1}{6}\, 
\big(F_{ijk}+F_{jki}+F_{kij}+F_{ikj}+F_{jik}+F_{kji}\big)\\
\label{TA}
{}_{\text{A}}F_{ijk}&=\frac{1}{6}\, 
\big(F_{ijk}+F_{jki}+F_{kij}-F_{ikj}-F_{jik}-F_{kji}\big)\\
\label{TPB}
{}_{\bar{\text{P}}}F_{ijk}&=\frac{1}{3}\, 
\big(F_{ijk}+F_{jik}-F_{kji}-F_{jki}\big)\\
\label{TP}
{}_{{\text{P}}}F_{ijk}&=\frac{1}{3}\, 
\big(F_{ijk}+F_{kji}-F_{jik}-F_{kij}\big)\,.
\end{align}
The four tensor pieces~(\ref{TS})--(\ref{TP}) are irreducible with respect to
the $n$-dimensional general linear group $GL(n)$ ($n>2$).
The tensor ${}_{\text{S}}F_{ijk}$ is a completely symmetric tensor
of rank three corresponding to the pattern (3), 
${}_{\text{A}}F_{ijk}$ is a completely antisymmetric tensor
of rank three corresponding to the pattern (1,1,1).
Since there are two standard tableaux for the pattern (2,1), two irreducible
tensors exist for this pattern, namely
${}_{\bar{\text{P}}}F_{ijk}$ which is symmetric in $i$ and $j$ and
antisymmetric in $i$ and $k$, and
${}_{{\text{P}}}F_{ijk}$ which is symmetric in $i$ and $k$ and antisymmetric in 
$i$ and $j$.~\footnote{Although the correct symmetry operators (Eqs.~(4.9) and (4.10) in
  \citep{Toupin62}) 
are used by~\citet{Toupin62}, 
the tensors ${}_{\text{A}}~\!\!a_{pqr}$ and ${}_{\bar{\text{P}}}a_{pqr}$
in Eq.~(4.11) in \citep{Toupin62} contain misprints. 
Moreover, the tensor decomposition based on Young tableaux given in this paper 
is in agreement with the decomposition given by~\citet{Wade44},
namely: 
${}_{\text{S}}F_{ijk}\equiv S_{ijk}$, 
${}_{\text{A}}F_{ijk}\equiv Q_{ijk}$, 
${}_{\bar{\text{P}}}F_{ijk}\equiv Q^{(1)}_{ijk}$, 
${}_{{\text{P}}}F_{ijk}\equiv Q^{(2)}_{ijk}$.}
Note that the tensor pieces~(\ref{TS}), (\ref{TPB}), (\ref{TP}) are reducible with respect
to the group $O(n)$, since they are not traceless; only the totally
antisymmetric piece~(\ref{TA}) is already irreducible with respect to the group
$O(n)$, since it is traceless by construction.
The irreducible decomposition of the general tensor $F_{ijk}$ with respect to $O(n)$ is given in Appendix~\ref{appendixE}.

The tensor pieces 
${}_{\text{S}}F_{ijk}$ and ${}_{\bar{\text{P}}}F_{ijk}$
are the relevant ones in strain gradient elasticity (see below),
whereas the pieces
${}_{\text{A}}F_{ijk}$ and ${}_{{\text{P}}}F_{ijk}$ are the relevant 
ones in couple-stress elasticity (see~\citep{Toupin62}), 
in dislocation gauge theory (see~\citep{Lazar02,LA09})
and in the so-called relaxed micromorphic elasticity (see~\citep{Neff}).
Moreover, the decomposition of the dislocation density tensor in its three irreducible tensor pieces (called `tentor', `trator', `axitor') with respect to the three-dimensional orthogonal group 
is given by~\citet{Lazar02} and \citet{LA09}. 
The corresponding decomposition of $F_{[ij]k}$ is given in Appendix~\ref{appendixD}.

\subsection{Decomposition of the strain gradient tensor}
Now we construct the $O(n)$-irreducible pieces of the strain gradient tensor 
$\eta_{ijk}$ using the method of Young tableaux.
The strain gradient tensor, by definition, is symmetric in its first
pair of indices. Thus, it is not a general tensor of rank three ($\eta_{ijk}\equiv F_{(ij)k}$).
%%This simplifies the whole procedure.

For the strain gradient tensor there are only two non-vanishing Young tableaux if $n>2$, namely 
\begin{eqnarray}\label{tab2}
 \Yvcentermath1\young({i}{j})\otimes\young(k)
 &=&\Yvcentermath1 \alpha\,\young({i}{j}{k})
 \Yvcentermath1 + \beta\,\young({i}{j},{k}) 
%\end{align}
\end{eqnarray}
\begin{align}
\label{deco-YT}
\eta_{(ij)k}&=\eta_{(ijk)}+\frac{4}{3}\, \eta_{(j[i)k]}\,
\end{align}
%%where $\alpha$ is a normalization factor.
%%%% $\alpha=1/12$.
with the two tensor pieces
\begin{align}
\label{eta-1-gl}
{}_{\text{S}}\eta_{ijk}:=
\eta_{(ijk)}&=\frac{1}{3}\big(\eta_{ijk}+\eta_{jki}+\eta_{kij}\big)\\
\label{eta-2-gl}
{}_{\bar{\text{P}}}\eta_{ijk}:=
\frac{4}{3}\, \eta_{(j[i)k]}&=
\frac{1}{3}\big(2 \eta_{ijk}-\eta_{jki}-\eta_{kij}\big)
\end{align}
which are irreducible with respect to $GL(n)$ and reducible with respect to $O(n)$.

In order to construct the $O(n)$-irreducible pieces of $\eta_{ijk}$, 
we split the strain gradient tensor $\eta_{ijk}$ into two parts
\begin{align}
\label{eta-deco}
\eta_{ijk}={{\eta\quer}}{}_{\!\!ijk} +\frac{1}{n}\,\delta_{ij}\, \eta_{\cdot\cdot k}
\end{align}
namely the {\it shear gradient tensor}
\begin{align}
%%%{\not\!\eta}_{ijk}=
{{\eta\quer}}_{\!\!ijk} :=
\eta_{ij k}-\frac{1}{n}\,\delta_{ij}\, \eta_{\cdot\cdot k}
={\varepsilon\quer}_{\!\! ij,k}
%%%\, ,\qquad{{\eta\quer}}{}_{\cdot\cdot  k}=0
\end{align}
with the two traces 
\begin{align}
&{{\eta\quer}}{}_{\!\!\cdot\cdot k}=0\\
%%\end{align}
%%and 
%%%\qquad\text{and}\qquad
%% \begin{align}
 \label{eta-q-vec}
 &{\eta\quer}_{\!\!i\cdot\cdot}=\eta_{i\cdot\cdot }-\frac{1}{n}\, 
\eta_{\cdot\cdot i}={\varepsilon\quer}_{\!\!ij,j}
\end{align}
and the {\it dilatation gradient vector}
\begin{align}
\eta_{\cdot\cdot k}:=\eta_{iik}=\delta_{ij}\eta_{ijk}
=\varepsilon_{\cdot\cdot,k}\,
\end{align}
which is nothing but the gradient of the dilatation $\varepsilon_{\cdot\cdot}$.
Eq.~(\ref{eta-q-vec}) is the non-vanishing trace of the shear gradient tensor.
In this formulation, 
the vector ${\eta\quer}_{\!\!i\cdot\cdot}$ in Eq.~(\ref{eta-q-vec}) 
is the {\it vector part of the shear gradient tensor}
and identically the {\it divergence of the shear tensor}.
Therefore, we call this vector part~(\ref{eta-q-vec}) the {\it shear divergence vector}. 
%%since it is the non-vanishing trace of the shear gradient tensor.
The dilatation gradient vector cannot be reduced any further, since it is
already an irreducible piece, and the application of the Young tableau method
to the shear gradient tensor together with the taking of traces yields three further
irreducible pieces, provided $n>2$.

Therefore, 
the strain gradient tensor $\eta_{ijk}$ can be decomposed into four
irreducible pieces
under the orthogonal group $O(n)$
according to 
%The irreducible parts of $\eta_{ijk}$ are given by
\begin{align}
&\eta_{ijk}={}^{(1)}\eta_{ijk}+{}^{(2)}\eta_{ijk}+{}^{(3)}\eta_{ijk}+{}^{(4)}\eta_{ijk}
\end{align}
possessing the number of independent tensor components
\begin{align}
&\frac{n^2(n+1)}{2}=\frac{n(n-1)(n+4)}{6}+\frac{n(n^2-4)}{3}+n+n\,. 
\end{align}
These pieces are invariant under the action of $O(n)$ and $SO(n)$.  
The irreducible pieces are given in the following.

\subsubsection{The irreducible piece: ${}^{(1)}\eta_{ijk}$}

The {\it totally symmetric and traceless tensor of rank three} is 
characterized by the following Young tableau:
\\
\\
%\vspace*{3mm}
\unitlength0.5cm
\begin{picture}(30,1)
\linethickness{0.075mm}
%\put(1,0){
\young({i}{j}{k})
%\put(1,0){\framebox(1,1){$i$}}
%\put(2,0){\framebox(1,1){$j$}}
%\put(3,0){\framebox(1,1){$k$}}
\put(2,0){$\stackrel{\wedge}{=}$}
\put(4,0){$\relstack{ijk}{\cal S} \eta_{ijk}- \mathrm{traces}$}
\end{picture}
\\
\\
where 
$\relstack{ijk}{\cal S}$ means the total symmetrization in the indices $i,j,k$
and divided by $3!$. 
Since the Young symmetry and the trace subtraction are `commutative' for the
construction of $O(n)$-irreducible tensors~\citep{Hamer}, we obtain
\begin{align}
\label{T1}
{}^{(1)}\eta_{ijk}
&=
\eta_{(ijk)}-\mathrm{traces}
=\tl \eta_{(ijk)}\,.
\end{align}
Substituting the traceless tensor of rank three~(\ref{F-tl}) into
Eq.~(\ref{T1}) and doing some
algebra, we find 
\begin{align}
\label{T1-1}
{}^{(1)}\eta_{ijk}&=
\frac{1}{3}\Big(\eta_{ijk}+\eta_{jki}+\eta_{kij}
-\frac{1}{n+2}\Big(\delta_{ij}\big[2\eta_{k\cdot\cdot}+\eta_{\cdot\cdot k}\big]
+\delta_{ik}\big[2\eta_{j\cdot\cdot }+\eta_{\cdot\cdot j}\big]
+\delta_{jk}\big[2\eta_{i\cdot\cdot}+\eta_{\cdot\cdot i}\big]\Big)
\Big)\,.
\end{align}
It is remarkable that the irreducible piece ${}^{(1)}\eta_{ijk}$ can be equivalently expressed in terms of the shear gradient tensor and the non-vanishing vector trace of the shear gradient tensor, which is the shear divergence vector, according to  
\begin{align}
\label{T1-2}
{}^{(1)}\eta_{ijk}&=
\frac{1}{3}\Big({\eta\quer}_{\!\!ijk}+{\eta\quer}_{\!\!jki}+{\eta\quer}_{\!\!kij}
-\frac{2}{n+2}\Big(\delta_{ij}\,{\eta\quer}_{\!\!k\cdot\cdot}
+\delta_{ik}\,{\eta\quer}_{\!\!j\cdot\cdot }
+\delta_{jk}\,{\eta\quer}_{\!\!i\cdot\cdot}\Big)
\Big)\,. 
\end{align}
Thus, the piece~(\ref{T1-2})
is the {\it totally symmetric and traceless shear gradient tensor 
of Young pattern~(3)} (traceless tensor of symmetry~(3)).

\subsubsection{The irreducible piece: ${}^{(2)}\eta_{ijk}$}
Another irreducible tensor part of $\eta_{ijk}$ is characterized by 
a traceless tensor of rank three having the following Young symmetry:
\\
\\
\\
\unitlength0.5cm
\begin{picture}(30,1)
\linethickness{0.075mm}
\young({i}{j},{k}) 
%\put(1,-1){\framebox(1,1){$k$}}
%\put(1,0){\framebox(1,1){$i$}}
%\put(2,0){\framebox(1,1){$j$}}
\put(2,1){$\stackrel{\wedge}{=}$}
\put(4,1)
{${\hbox{\Large$\frac{4}{3}$}}\,
\relstack{ki}{\cal A}\,
\relstack{ij}{\cal S}
\eta_{ijk}
-\mathrm{traces}$}
\end{picture}
\\ 
\\
where 
$\relstack{ij}{\cal S}$ means the symmetrization in
the indices $i,j$ and divided by $2$, and 
$\relstack{ki}{\cal A}$ means the antisymmetrization in the indices $k,i$
and divided by $2$. The factor $4/3$ is the normalization
factor for the tensor decomposition~(\ref{deco-YT}). 
Using that the Young symmetry and the trace subtraction are `commutative'
operations for the
construction of $O(n)$-irreducible tensors~\citep{Hamer}, we get
\begin{align}
\label{T2}
{}^{(2)}\eta_{ijk}
&=
\frac{4}{3}\, \eta_{(j[i)k]}-\mathrm{traces}
=\frac{4}{3}\, \tl \eta_{(j[i)k]}\,. 
\end{align}
Therefore, the tensor  ${}^{(2)}\eta_{ijk}$ is symmetric in the indices $i$ and $j$, 
and antisymmetric in the indices $i$ and $k$.
Substituting the traceless tensor of rank three~(\ref{F-tl}) into
Eq.~(\ref{T2}) and doing some algebra, we obtain
\begin{align}
\label{T2-1}
{}^{(2)}\eta_{ijk}&=
\frac{1}{3}\Big(2 \eta_{ijk}-\eta_{jki}-\eta_{kij}
+\frac{1}{n-1}\Big(2\delta_{ij}\big[\eta_{k\cdot\cdot}-\eta_{\cdot\cdot k}\big]
-\delta_{ik}\big[\eta_{j\cdot\cdot}-\eta_{\cdot\cdot j}\big]
-\delta_{jk}\big[\eta_{i\cdot\cdot}-\eta_{\cdot\cdot i}\big]\Big)
\Big)\,.
\end{align}
Note that the irreducible piece ${}^{(2)}\eta_{ijk}$ can be equivalently
expressed in terms of the shear gradient tensor 
and 
%%the non-vanishing vector trace of the shear gradient tensor, namely 
the shear divergence vector  according to  
\begin{align}
\label{T2-2}
{}^{(2)}\eta_{ijk}&=
\frac{1}{3}\Big(2\,{\eta\quer}_{\!\!ijk}-{\eta\quer}_{\!\!jki}-{\eta\quer}_{\!\!kij}
+\frac{1}{n-1}\Big(2\delta_{ij}\,{\eta\quer}_{\!\!k\cdot\cdot}
-\delta_{ik}\,{\eta\quer}_{\!\!j\cdot\cdot}
-\delta_{jk}\,{\eta\quer}_{\!\!i\cdot\cdot}\Big)
\Big)\,.
\end{align}
Thus, the piece~(\ref{T2-2})
is the {\it traceless shear gradient tensor with mixed symmetry of 
Young pattern~(2,1)} 
(traceless tensor of symmetry~(2,1)).
%%%(symmetric in the indices $i$ and $j$ and antisymmetric in the indices $i$ and $k$).

\subsubsection{The irreducible pieces: ${}^{(3)}\eta_{ijk}$ and ${}^{(4)}\eta_{ijk}$}
There are two irreducible vector pieces corresponding to the two canonical traces,
namely the dilatation gradient vector $\eta_{\cdot\cdot k}$ and 
the shear divergence vector ${\eta\quer}_{\!\!i\cdot\cdot}$.
The dilatation gradient vector gives the irreducible piece  ${}^{(4)}\eta_{ijk}$:
\begin{align}
\label{T4-1}
{}^{(4)}\eta_{ijk}&=\frac{1}{n}\,\delta_{ij}\, \eta_{\cdot\cdot k}\,.
\end{align}
Using the irreducible decomposition,
the {\it condition of incompressibility} can be stated as
\begin{align}
\label{incompr}
\varepsilon_{\cdot\cdot}=0\,,\qquad\eta_{\cdot\cdot k}=0\,, \qquad {}^{(4)}\eta_{ijk}=0\,.
\end{align}

The other irreducible vector piece is defined by 
\begin{align}
\label{T3}
{}^{(3)}\eta_{ijk}= \eta_{ijk}-{}^{(1)}\eta_{ijk}-{}^{(2)}\eta_{ijk}-{}^{(4)}\eta_{ijk}&
\end{align}
and reads
\begin{align}
\label{T3-1}
{}^{(3)}\eta_{ijk}&=\frac{n}{(n+2)(n-1)}
\Big(\delta_{ik}\Big[\eta_{j\cdot\cdot}-\frac{1}{n}\,\eta_{\cdot\cdot j}\Big]
+\delta_{jk}\Big[\eta_{i\cdot\cdot}-\frac{1}{n}\,\eta_{\cdot\cdot i}\Big]
-\frac{2}{n}\,\delta_{ij}\Big[\eta_{k\cdot\cdot}-\frac{1}{n}\,\eta_{\cdot\cdot k}\Big]\Big)
\Big)\,.
\end{align}
The irreducible piece ${}^{(3)}\eta_{ijk}$ can be completely expressed in terms of the
 shear divergence vector according to  
\begin{align}
\label{T3-2}
{}^{(3)}\eta_{ijk}&=\frac{n}{(n+2)(n-1)}
\Big(\delta_{ik}\,{\eta\quer}_{\!\!j\cdot\cdot}+\delta_{jk}\,{\eta\quer}_{\!\!i\cdot\cdot}
-\frac{2}{n}\,\delta_{ij}\, {\eta\quer}_{\!\!k\cdot\cdot}\Big)\,.
\end{align}
It can be seen that the irreducible piece given in Eq.~(\ref{T3-2})
is the sum of the traces subtracted out in the pieces
${}^{(1)}\eta_{ijk}$ in Eq.~(\ref{T1-2}) and
${}^{(2)}\eta_{ijk}$ in Eq.~(\ref{T2-2}).

\subsection{Properties of the irreducible pieces} 

First of all, we observe that the {\it presented decomposition of the strain gradient tensor 
$\eta_{ijk}$ is canonical and unique}, since we started from the 
initial decomposition~(\ref{eta-deco}) into the shear gradient 
tensor and the dilatation gradient vector, 
whereas the latter is already irreducible. 
The initial decomposition~(\ref{eta-deco}) has ensured that the irreducible
subspaces which we have obtained are uniquely defined.
The unique decomposition of the shear gradient tensor into its three irreducible
pieces is given by
\begin{align}
&{\eta\quer}_{\!\!ijk}={}^{(1)}\eta_{ijk}+{}^{(2)}\eta_{ijk}+{}^{(3)}\eta_{ijk}
\end{align}
which are given by Eqs.~(\ref{T1-2}), (\ref{T2-2}) and (\ref{T3-2}).

There are {\it two types of vanishing traces}
\begin{align}
&{}^{(1)}\eta_{\cdot\cdot k}={}^{(2)}\eta_{\cdot\cdot k}={}^{(3)}\eta_{\cdot\cdot k}=0\\
&{}^{(1)}\eta_{i\cdot\cdot}={}^{(2)}\eta_{i\cdot\cdot}=0
\end{align}
and the {\it Young symmetries of those parts vanish} are given by
\begin{align}
&{}^{(1)}\eta_{i[jk]}=0\,,\qquad {}^{(2)}\eta_{(ijk)}=0\,.
\end{align}
There are {\it two types of non-vanishing traces}
\begin{align}
\label{T3-tr}
{}^{(3)}\eta_{i\cdot\cdot}&={\eta\quer}_{\!\! i\cdot\cdot}\\
\label{T4-tr2}
{}^{(4)}\eta_{i \cdot\cdot}&=\frac{1}{n}\,\eta_{\cdot\cdot i}\\\
\label{T4-tr1}
{}^{(4)}\eta_{\cdot\cdot k}&=\eta_{\cdot\cdot k}\,.
\end{align}

The four irreducible components defined by Eqs.~(\ref{T1-2}), (\ref{T2-2}), (\ref{T4-1}) and (\ref{T3-2}) 
have an interesting {\it orthogonality property}
\begin{align}
&{}^{(I)}\eta_{ijk}\,{}^{(J)}\eta_{ijk}=0\,,\qquad I\neq J\\
&{}^{(I)}\eta_{ijk}\,{}^{(J)}\eta_{ijk}\neq 0\,,\qquad I = J
\end{align}
so that 
\begin{align}
\eta_{ijk}\eta_{ijk}=
\sum_{I=1}^{4} {}^{(I)}\eta_{ijk}{}^{(I)}\eta_{ijk}\,.
\end{align}
On the other hand, it holds one {\it cross term} relation
between ${}^{(3)}\eta_{ijk}$ and ${}^{(4)}\eta_{kji}$
\begin{align}
 {}^{(3)}\eta_{ijk}{}^{(4)}\eta_{kji}=
\frac{1}{n}\  {\eta\quer}_{\!\!i\cdot\cdot} \eta_{\cdot\cdot i}
%%%%\neq 0\,
\end{align}
describing the coupling between the shear divergence vector 
${\eta\quer}_{\!\!i\cdot\cdot}$ and the dilatation gradient vector
$\eta_{\cdot\cdot i}$.

\subsection{Three dimensions: $n=3$}

In three dimensions, 
the irreducible pieces of $\eta_{ijk}$ are given 
by\footnote{In three dimensions, the irreducible tensor pieces
${}^{(1)}\eta_{ijk}$, ${}^{(2)}\eta_{ijk}$, 
${}^{(3)}\eta_{ijk}$, ${}^{(4)}\eta_{ijk}$ have a spin $j=3,2,1,1$, respectively.
The spin $j$ is the highest weight which characterizes irreducible
representations up to equivalence (see, e.g.,~\citep{Sexl}).
}
\begin{align}
\eta_{ijk}&={}^{(1)}\eta_{ijk}+{}^{(2)}\eta_{ijk}+{}^{(3)}\eta_{ijk}+{}^{(4)}\eta_{ijk}\\
%\end{align}
%\begin{align}
18&=7+5+3+3
\end{align}
where\footnote{Note that the tensor decompositions given in~\citep{Auffray13,MB} 
are not canonical since the tensors $S_1(\bm a)_{ijk}$ and $S_4(\bm a)_{ijk}$ 
in~\citep{MB} and the tensors $T(\mathrm{V}^{\nabla\mathrm{str}})_{ijk}$, and 
$T(\mathrm{V}^{\nabla\mathrm{rot}})_{ijk}$ in \citep{Auffray13}
are arbitrary combinations of the irreducible 
pieces ${}^{(3)}\eta_{ijk}$ and ${}^{(4)}\eta_{ijk}$.}
\begin{align}
\label{T1-2-3d}
{}^{(1)}\eta_{ijk}&=
\frac{1}{3}\Big({\eta\quer}_{\!\!ijk}+{\eta\quer}_{\!\!jki}+{\eta\quer}_{\!\!kij}
-\frac{2}{5}\Big(\delta_{ij}\,{\eta\quer}_{\!\!k\cdot\cdot}
+\delta_{ik}\,{\eta\quer}_{\!\!j\cdot\cdot }
+\delta_{jk}\,{\eta\quer}_{\!\!i\cdot\cdot}\Big)\Big)\\
%%\end{align}
%%\begin{align}
\label{T2-2-3d}
{}^{(2)}\eta_{ijk}&=
\frac{1}{3}\Big(2\,{\eta\quer}_{\!\!ijk}-{\eta\quer}_{\!\!jki}-{\eta\quer}_{\!\!kij}
+\frac{1}{2}\Big(2\delta_{ij}\,{\eta\quer}_{\!\!k\cdot\cdot}
-\delta_{ik}\,{\eta\quer}_{\!\!j\cdot\cdot}
-\delta_{jk}\,{\eta\quer}_{\!\!i\cdot\cdot}\Big)
\Big)\\
%%\end{align}
%%\begin{align}
\label{T3-2-3d}
{}^{(3)}\eta_{ijk}&=\frac{3}{10}
\Big(\delta_{ik}\,{\eta\quer}_{\!\!j\cdot\cdot}+\delta_{jk}\,{\eta\quer}_{\!\!i\cdot\cdot}
-\frac{2}{3}\,\delta_{ij}\, {\eta\quer}_{\!\!k\cdot\cdot}\Big)\\
%%\end{align}
%%\begin{align}
\label{T4-1-3d}
{}^{(4)}\eta_{ijk}&=\frac{1}{3}\,\delta_{ij}\, \eta_{\cdot\cdot k}\,.
\end{align}

\subsection{Two dimensions: $n=2$}

In two dimensions, ${}^{(2)}\eta_{ijk}$ drops out leaving three irreducible
pieces due to the theorem that 
traceless tensors corresponding to Young tableaux in which the sum of the
lengths of the first
two columns exceeds $n=2$ must be identically zero~\citep{Hamer}.

Thus, the irreducible pieces of $\eta_{ijk}$ are given by
%The irreducible parts of $\eta_{ijk}$ are given by
\begin{align}
\eta_{ijk}&={}^{(1)}\eta_{ijk}+{}^{(3)}\eta_{ijk}+{}^{(4)}\eta_{ijk}\\
%\end{align}
%\begin{align}
6&=2+2+2
\end{align}
where\footnote{The three irreducible pieces ${}^{(1)}\eta_{ijk}$,
${}^{(3)}\eta_{ijk}$, and ${}^{(4)}\eta_{ijk}$
correspond to the three tensors $H_{(ijk)}$, 
$T(\mathrm{V}^{\nabla\mathrm{dev}})_{ijk}$, and 
$T(\mathrm{V}^{\nabla\mathrm{sph}})_{ijk}$ given in~\citep{Auffray15}.}
\begin{align}
\label{T1-2-2d}
{}^{(1)}\eta_{ijk}&=
\frac{1}{3}\Big({\eta\quer}_{\!\!ijk}+{\eta\quer}_{\!\!jki}+{\eta\quer}_{\!\!kij}
-\frac{1}{2}\Big(\delta_{ij}\,{\eta\quer}_{\!\!k\cdot\cdot}
+\delta_{ik}\,{\eta\quer}_{\!\!j\cdot\cdot }
+\delta_{jk}\,{\eta\quer}_{\!\!i\cdot\cdot}\Big)\Big)\\
\label{T3-2-2d}
{}^{(3)}\eta_{ijk}&=\frac{1}{2}
\Big(\delta_{ik}\,{\eta\quer}_{\!\!j\cdot\cdot}+\delta_{jk}\,{\eta\quer}_{\!\!i\cdot\cdot}
-\delta_{ij}\, {\eta\quer}_{\!\!k\cdot\cdot}\Big)\\
%%\end{align}
%%\begin{align}
\label{T4-1-2d}
{}^{(4)}\eta_{ijk}&=\frac{1}{2}\,\delta_{ij}\, \eta_{\cdot\cdot k}\,.
\end{align}

\section{The strain energy density in terms of  irreducible strain and irreducible strain gradient tensors}

The most general quadratic form of 
the {\it strain energy density of isotropic strain gradient elasticity} given in terms of the irreducible strain and 
irreducible strain gradient tensors is given by 
\begin{align}
\label{W-irrep}
W=\frac{1}{2}\, \varepsilon_{ij}\left(\sum_{I=1}^2 c_I\,  {}^{(I)}\varepsilon_{ij}\right)
%%%\Big(c_1 \ {\varepsilon\quer}_{\!\!ij}+\frac{c_2}{n}\,\delta_{ij}\, \varepsilon_{\cdot\cdot}\Big)
+\frac{1}{2}\, \eta_{ijk} \left(\sum_{I=1}^4 b_I\,  {}^{(I)}\eta_{ijk}\right)
+b_5\,  {}^{(3)}\eta_{ijk}{}^{(4)}\eta_{kji}\,.
\end{align}
Here, $c_1$ and $c_2$ are `{\it elastic constants}' and $b_1,\dots, b_5$ are 
five `{\it irreducible gradient parameters}' due to the irreducible pieces.  
The gradient parameters $b_1,\dots, b_4$ are the `coupling constants'
for the four irreducible pieces of the strain gradient tensor 
${}^{(1)}\eta_{ijk},\dots, {}^{(4)}\eta_{ijk}$. 
In particular, $b_4$ is the `coupling constant' for the dilatation  gradient
vector $\eta_{\cdot\cdot k}$,
and $b_1, b_2, b_3$ are the `coupling constants' for the 
three irreducible pieces of the shear gradient tensor
${}^{(1)}\eta_{ijk},{}^{(2)}\eta_{ijk}, {}^{(3)}\eta_{ijk}$, respectively. 
Note the peculiar cross term with gradient parameter $b_5$
\begin{align}
{}^{(3)}\eta_{ijk}{}^{(4)}\eta_{kji}=
\frac{1}{n}\  {\eta\quer}_{\!\!i\cdot\cdot} \eta_{\cdot\cdot i}
=
\frac{1}{n}\,\eta_{i\cdot\cdot} \eta_{\cdot\cdot i}
-\frac{1}{n^2}\, \eta_{\cdot\cdot i}\eta_{\cdot\cdot i}\,.
\end{align}
In fact, the gradient parameter $b_5$ is the `coupling constant' between
the shear divergence vector 
${\eta\quer}_{\!\!i\cdot\cdot}$ and the dilatation gradient vector
$\eta_{\cdot\cdot i}$. 
Also, notice that since for  $n=2$ the piece ${}^{(2)}\eta_{ijk}$ drops out, there are
only four gradient parameters: $b_1, b_3, b_4, b_5$ in two-dimensional 
isotropic strain gradient elasticity.

We can connect the constants $c_1$ and $c_2$ with 
the moduli known from elasticity theory
\begin{align}
c_1=2\mu\,, \qquad c_2=n\, K\,,
\qquad K=\lambda+\frac{2}{n}\, \mu
\end{align}
where $\mu$ and $\lambda$ are the {\it Lam\'e constants} 
and $K$ is the {\it modulus of compression} or {\it bulk modulus} (see, e.g., \citep{LL}) 
for $n$-dimensions.

Using Eq.~({\ref{W-irrep}), we define the conjugate quantities of isotropic strain gradient elasticity, namely
the {\it Cauchy stress tensor} is given by
\begin{align}
\label{CS-irrep}
\sigma_{ij}:=\frac{\pd W}{\pd \varepsilon_{ij}}
=\sum_{I=1}^2 c_I\,  {}^{(I)}\varepsilon_{ij}
=c_1\ {\varepsilon\quer}_{\!\!ij} +\frac{c_2}{n}\, \delta_{ij} \varepsilon_{\cdot\cdot}
\end{align}
and the {\it double stress tensor} is given by
\begin{align}
\label{DS-irrep}
\tau_{ijk}:=\frac{\pd W}{\pd \eta_{ijk}}=\sum_{I=1}^4 b_I\,  {}^{(I)}\eta_{ijk}
+ \frac{b_5}{2n}\, \big(\delta_{ik}\eta_{llj}+\delta_{jk}\eta_{lli}+2\delta_{ij}\eta_{kll}\big)
           %%%+\nonumber\\	   &\quad
	   -\frac{2 b_5 }{n^2}\,\delta_{ij}\eta_{llk}\,. 
\end{align}

By means of setting $b_5=0$, we `switch off' the coupling between the 
shear divergence vector and the dilatation gradient vector and 
obtain the double stress tensor in terms 
of the four irreducible pieces and 
the corresponding four gradient parameters 
\begin{align}
\label{DS-irrep2}
\tau_{ijk}=\sum_{I=1}^4 b_I\,  {}^{(I)}\eta_{ijk}
\,. 
\end{align}

In strain gradient elasticity, 
the equilibrium condition  reads
\begin{align}
\label{EC}
\sigma_{ij,j}-\tau_{ijk,kj}+f_i=0
\end{align}
where $f_i$ is the body force density vector.

\section{Relation to Mindlin's strain gradient elasticity of form II}

\subsection{The general case: $n\ge 3$}

The strain energy density of 
{\it Mindlin's isotropic strain gradient elasticity theory of form II} 
reads~\citep{Mindlin64} (see also~\citep{Mindlin68,Shodja}) 
\begin{align}
\label{W-M}
W=\frac{1}{2}\, \lambda\, \varepsilon_{ii}\varepsilon_{jj}+\mu\,\varepsilon_{ij}\varepsilon_{ij}
+a_1\,\eta_{iik}\eta_{kjj}
+a_2\, \eta_{iik}\eta_{jjk}
+a_3\,\eta_{ijj}\eta_{ikk}
+a_4\,\eta_{ijk}\eta_{ijk}
+a_5\,\eta_{ijk}\eta_{kji}\,
\end{align}
where $a_1$, $a_2$, $a_3$, $a_4$, $a_5$ are the {\it Mindlin gradient 
parameters} in Mindlin's first strain gradient elasticity theory.

Now the Cauchy stress tensor reads
\begin{align}
\label{CS-M}
\sigma_{ij}=
%%=\frac{\pd W}{\pd \varepsilon_{ij}}=
2 \mu\, \varepsilon_{ij} +\lambda\, \delta_{ij} \varepsilon_{\cdot\cdot}
\end{align}
and the double stress tensor is given by
\begin{align}
\label{DS-M}
\tau_{ijk}%%%&=\frac{\pd W}{\pd\big( \eta_{ijk}}
          =\frac{a_1}{2} \big(\delta_{ik}\eta_{llj}+\delta_{jk}\eta_{lli}+2\delta_{ij}\eta_{kll}\big)
           %%%+\nonumber\\	   &\quad
	   +2 a_2 \delta_{ij}\eta_{llk}
	    +a_3\big(\delta_{ik}\eta_{jll}+\delta_{jk}\eta_{ill}\big)
	   +2 a_4 \eta_{ijk}
	   +a_5 \big(\eta_{jki}+\eta_{kij}\big)\, .
\end{align}

Substituting the irreducible pieces~(\ref{T1-1}), (\ref{T2-1}), (\ref{T4-1}),
and (\ref{T3-1}) into Eq.~(\ref{DS-irrep}), and comparing now with 
Eq.~(\ref{DS-M}), we obtain the relation
between Mindlin's gradient parameters and 
the `irreducible gradient parameters' 
\begin{align}
\label{rel-a1}
a_1&=
-\frac{2 b_1}{3(n+2)}+\frac{ 2 b_2}{3(n-1)}
-\frac{2 b_3}{(n+2)(n-1)}+\frac{1}{n} \, b_5\\
\label{rel-a2}
a_2&=-\frac{b_1}{6(n+2)}-\frac{b_2}{3(n-1)}
+\frac{b_3}{(n+2)(n-1)n}
+\frac{1}{2n} \, b_4 -\frac{1}{n^2} \, b_5\\
\label{rel-a3}
a_3&=-\frac{2 b_1}{3(n+2)}-\frac{ b_2}{3(n-1)}
+\frac{n b_3}{(n+2)(n-1)} \\
\label{rel-a4}
a_4&=\frac{1}{6}\, b_1+\frac{1}{3}\, b_2\\
\label{rel-a5}
a_5&=\frac{1}{3}\, b_1-\frac{1}{3}\, b_2
\end{align}
and the inverse relations between the `irreducible gradient parameters'
and  Mindlin's gradient parameters 
\begin{align}
\label{rel-b1}
b_1&= 2 a_4+2 a_5\\
b_2&=2 a_4 -a_5\\ 
b_3&= \frac{(n+2)(n-1) a_3+2n a_4+ (n-2)a_5}{n}\\
b_4&= 2\, \frac{n a_1 +n^2 a_2 +a_3+na_4+a_5}{n}\\
\label{rel-b5}
b_5& = n a_1+2 a_3+2 a_5\,.
\end{align}

If $b_5=0$ (no coupling between the shear divergence vector and the dilatation
gradient vector), then one parameter of the five Mindlin gradient parameters
can be eliminated using the constraint obtained from Eq.~(\ref{rel-b5})
\begin{align}
\label{constr}
n a_1+2 a_3+2 a_5=0\,.
\end{align}

\subsection{The case: $n=2$}
Because for  $n=2$ the piece ${}^{(2)}\eta_{ijk}$ drops out,  
there are only four gradient parameters: $b_1, b_3, b_4, b_5$
and therefore
\begin{align}
\label{redl-2d}
b_2=0:\quad \Longrightarrow\qquad a_5=2 a_4\,.
\end{align}
Using Eq.~(\ref{redl-2d}), 
the strain energy density~(\ref{W-M})
of {\it Mindlin's isotropic strain gradient elasticity theory of form II} 
reads in {\it two dimensions} 
\begin{align}
\label{W-M-2d}
W=\frac{1}{2}\,\lambda\, \varepsilon_{ii}\varepsilon_{jj}+\mu\,\varepsilon_{ij}\varepsilon_{ij}
+a_1\,\eta_{iik}\eta_{kjj}
+a_2\, \eta_{iik}\eta_{jjk}
+a_3\,\eta_{ijj}\eta_{ikk}
+a_4 \big( \eta_{ijk}\eta_{ijk}+ 2\eta_{ijk}\eta_{kji}\big)\,
\end{align}
and the double stress tensor~(\ref{DS-M}) is given by
\begin{align}
\label{DS-M-2d}
\tau_{ijk}%%%&=\frac{\pd W}{\pd\big( \eta_{ijk}}
          =\frac{a_1}{2} \big(\delta_{ik}\eta_{llj}+\delta_{jk}\eta_{lli}+2\delta_{ij}\eta_{kll}\big)
           %%%+\nonumber\\	   &\quad
	   +2 a_2 \delta_{ij}\eta_{llk}
	    +a_3\big(\delta_{ik}\eta_{jll}+\delta_{jk}\eta_{ill}\big)
	   +2 a_4\big( \eta_{ijk}+\eta_{jki}+\eta_{kij}\big)\, 
\end{align}
in terms of four (two-dimensional) Mindlin gradient parameters
$a_1,\dots,a_4$.

If $b_5=0$, then one parameter of the four Mindlin gradient parameter 
can be eliminated by substituting Eq.~(\ref{rel-b5}) into Eq.~(\ref{redl-2d})
\begin{align}
\label{constr-2d}
b_5=0:\quad \Longrightarrow\qquad a_4=-\frac{a_1+a_3}{2}\,.
\end{align}

\section{Relation to strain gradient elasticity of Helmholtz type}
For $n=3$,
a simplified and useful version of strain gradient elasticity
is called {\it strain gradient elasticity of Helmholtz type} 
which is a particular case of Mindlin's strain gradient elasticity of form II
(see, e.g., \citep{LM05,LMA05,Lazar13,Lazar14}). 
The connection between 
the strain energy density of isotropic strain gradient
elasticity given in terms of the irreducible strain  and 
the  strain energy density of strain gradient elasticity of Helmholtz type
is given by the particular choice of the gradient parameters
\begin{align}
\label{rel-Helm-b}
b_1=b_2=b_3=\ell^2 c_1\, ,\qquad b_4=\ell^2 c_2\,, \qquad b_5=0
\end{align}
where $\ell$ is a length scale parameter,
and equivalently in terms of Mindlin's gradient parameters
\begin{align}
\label{rel-Helm-a}
a_1=a_3=a_5=0\,,\qquad a_2=-\frac{\ell^2 c_1}{6}+\frac{\ell^2 c_2}{6}
\, ,\qquad a_4=\frac{\ell^2 c_1}{2}\,.
\end{align}
Then the double stress tensor~(\ref{DS-irrep}) reads 
\begin{align}
\label{DS-Helm}
\tau_{ijk}=%%:=\frac{\pd W}{\pd \eta_{ijk}}=
\ell^2 \Big( c_1\ {\eta\quer}_{\!\!ijk} +\frac{c_2}{3}\, \delta_{ij}\,
\eta_{\cdot\cdot k}\Big)
=\ell^2 \big( 2\mu\, {\eta}_{ijk} +\lambda\, \delta_{ij}\,\eta_{\cdot\cdot k}\big)
\,. 
\end{align}

Thus, gradient elasticity of Helmholtz type is based on the 
initial decomposition~(\ref{eta-deco}) into the shear gradient 
tensor ${\eta\quer}_{\!\!ijk}$  and the dilatation gradient vector 
$\eta_{\cdot\cdot k}$.

\section{Conclusions}

The irreducible decompositions presented here are of interest in themselves, both from the mechanical and group-theoretical point of view. 
One main motivation for the present work, based on the well proven relevance in strain gradient theories, is their use in setting
up the strain energy density and Lagrangian for isotropic strain gradient theories.

Using group theory, a canonical and unique tensor decomposition of the strain
gradient tensor into the {\it $O(n)$-irreducible pieces} has been given.
We have found that:
\begin{itemize}
\item
$n=3$: the strain gradient tensor can be decomposed into {\it four irreducible 
tensor pieces} with  {\it 7+5+3+3 independent components}
\item
$n=2$: the strain gradient tensor can be decomposed into {\it three irreducible
tensor pieces} with {\it 2+2+2 independent components}. 
\end{itemize}
The four irreducible pieces are built up from the three parts:
\begin{itemize}
\item
{\it shear gradient tensor} ${\eta\quer}_{\!\!ijk}$\,:
 ${}^{(1)}\eta_{ijk}$, ${}^{(2)}\eta_{ijk}$
\item
{\it shear divergence vector}
${\eta\quer}_{\!\!i\cdot\cdot}$\,:
 ${}^{(1)}\eta_{ijk}$, ${}^{(2)}\eta_{ijk}$,  ${}^{(3)}\eta_{ijk}$
\item
{\it dilatation gradient vector} $\eta_{\cdot\cdot k}$\,: 
 ${}^{(4)}\eta_{ijk}$\,.
\end{itemize}
The group-theoretical interpretation of the four irreducible pieces 
under $SO(3)$ is:
\begin{itemize}
\item
 ${}^{(1)}\eta_{ijk}$:
totally symmetric and traceless shear gradient tensor is spin-3 field 
\item
 ${}^{(2)}\eta_{ijk}$:
traceless shear gradient tensor of Young symmetry (2,1) is spin-2 field
\item
 ${}^{(3)}\eta_{ijk}$:
shear divergence vector in the 
subtracted traces of  ${}^{(1)}\eta_{ijk}$ and  ${}^{(2)}\eta_{ijk}$ is spin-1 field
\item
 ${}^{(4)}\eta_{ijk}$: gradient of the dilatation is spin-1 field.
\end{itemize}
As a straightforward consequence of the irreducible tensor decomposition,
we found the {\it number of gradient parameters:}
\begin{itemize}
\item
$n=3$: {\it five gradient parameters}
\item
$n=2$: {\it four gradient parameters}
\end{itemize}
in addition to the two Lam\'e constants.

Also we discussed the particular case, $b_5=0$,   
when the coupling between the shear divergence vector and the dilatation
gradient vector vanishes.

The presented irreducible decomposition of the strain gradient tensor can be
also used in strain gradient plasticity~(e.g.,~\citep{Fleck,Gao})
where usually a decomposition into three tensor pieces is used.

\section*{Acknowledgements}
The author gratefully acknowledges grants from the
Deutsche Forschungsgemeinschaft (Grant Nos.  La1974/3-1 and La1974/3-2).
In addition, useful remarks from Rainer Gl\"uge are gratefully acknowledged.

\begin{appendix}

\section{Young tableaux, Young operators and irreducible tensors}
\label{young}
\setcounter{equation}{0}
\renewcommand{\theequation}{\thesection.\arabic{equation}}

In this Appendix for the convenience of the reader
we collect some facts about Young tableaux and irreducible tensors
(see~\cite{BR,Boerner, Hamer,Weyl1,Weyl}). 

An irreducible representation $\Delta^{[m]}$ of the 
symmetric group $S_n$ is uniquely determined by the idempotent 
(normalized) {\it Young operator}
\begin{align}
\label{Young1}
{\cal Y}_{[m]} &= \frac{f_{[m]}}{n!}\,
{\cal Q}\, {\cal P}
\end{align}
with
\begin{align}
{\cal P} &= \sum_{p \in H_{[m]}} p
\qquad\quad \text{`symmetrizer'} \\
{\cal Q} &= \sum_{q \in V_{[m]}} \delta_q\,q
\qquad  \text{`antisymmetrizer'}
\end{align}
and
\begin{align}
\label{Young2}
{\cal Y}_{[m]}{\cal Y}_{[m']}
&=
\delta_{[m][m']}{\cal Y}_{[m]}
\end{align}
which is related to a Young tableau denoted by $[m]$.
Here ${\cal P}$ is the operator for {\it horizontal permutations} in the diagram 
and ${\cal Q}$ is the operator for {\it vertical permutations}.
Horizontal permutations $p$ are permutations which interchange only symbols 
(or indices) in the same row.
Vertical permutations $q$ interchange only symbols (or indices) in the same column
and $\delta_q$ is the parity of the permutation $q$. 

If
\begin{align}
{\underline m} = (m_1, m_2, \ldots m_r) 
\quad
{\rm with}
\quad
m_1 \geq m_2\geq \ldots \geq m_r\,,
\quad
%{\rm and}
%\quad
\sum^r_{i=1}\, m_i = n
\end{align}
defines a {\it Young pattern} (Fig.~\ref{Y-Rahmen}), then a {\em 
Young tableau} $[m]$ is obtained by putting in (without
repetition) the indices $i_1, \ldots i_n$ 
of a tensor  of rank $n$, $F_{i_1\dots i_n}$,
and $H_{[m]}$ and $V_{[m]}$ denotes 
their horizontal and vertical permutations with respect to
$[m]$.
A {\em standard tableau} is obtained when the indices
$i_1, \ldots, i_n$ are ordered lexicographically. There are
\begin{align}
\label{f}
f_{[m]} = n! \, 
\frac{\prod_{i<j} (l_i - l_j)}{\prod_{i=1}^r l_i!}
\quad
{\rm with}
\quad
l_i = m_i + r - i\,,
\quad
\sum_{[m]}f_{[m]}^2 = n!
\end{align}
different standard tableaux which correspond to $f_{[m]}$ 
different, but equivalent, irreducible representations of $S_n$
whose dimension is given also by $f_{[m]}$.
The (normalized) Young operators ${\cal Y}_{[m]}$ 
according to Eq.~(\ref{Young2}) project onto 
mutually orthogonal irreducible right ideals which provide irreducible 
representations of $S_n$.
%%%%%%%%%%%%%%%%%%%%%%%%%%%%%%%%%%%%%%%%%%%%
%%%%%%%%%%%%%%%%%%%
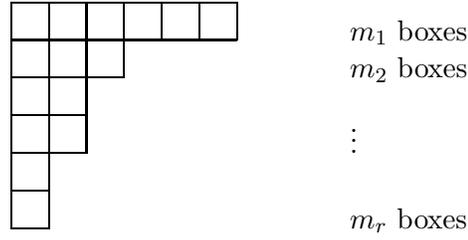
\begin{figure}[t]
\unitlength0.5cm
\begin{center}
\begin{picture}(15,5)
\linethickness{0.15mm}
\multiput(3,4)(1,0){5}{\line(0,1){1}}
\multiput(1,-1)(1,0){2}{\line(0,1){6}}
\multiput(3,1)(3,0){1}{\line(0,1){4}}
\multiput(4,3)(4,0){1}{\line(0,1){2}}
\put(1,5){\line(1,0){6}}
\put(1,4){\line(1,0){6}}
\put(1,3){\line(1,0){3}}
\put(1,2){\line(1,0){2}}
\put(1,1){\line(1,0){2}}
\put(1,0){\line(1,0){1}}
\put(1,-1){\line(1,0){1}}
\put(10,4){$m_1$ boxes}
\put(10,3){$m_2$ boxes}
\put(10,1){$\vdots$ }
\put(10,-1){$m_r$ boxes}
\end{picture}
\end{center}
\caption{\label{Y-Rahmen} Young pattern $\underline m$} 
\end{figure}
%%%%%%%%%%%%%%%%%%%%%%%%%%%%%%%%%%%%%%%

Thus, in order to obtain a tensor of rank $n$ 
with the symmetry described by a Young pattern 
${\underline m}$, we apply the Young operator ${\cal Y}_{[m]}$ to
the tensor $F_{i_1\dots i_n}$:
\begin{align}
\label{Fm}
F_{i_1  \ldots i_n}^{[m]}={\cal Y}_{[m]} 
F_{i_1  \ldots i_n}\,.
\end{align}
Consequently, 
the tensor will be 
symmetric in all the indices which appear in the same row
and antisymmetric in all the indices which appear in the same 
column. Any tensor component 
for which an index appears twice in the same column is equal to zero.
A tensor having a certain Young symmetry is irreducible with respect 
to the general linear group $GL(n)$.

Going from the group $GL(n)$ to the orthogonal group $O(n)$,  
only the completely antisymmetric tensors remain irreducible. 
The reason is that, because of the  definition 
of the orthogonal group,
$\delta_{ij} a_{ik} a_{jl} = \delta_{kl}, \;\forall {\bm a} \in O(n)$, 
the operation of taking the trace (contraction) of a tensor 
commutes with the orthogonal transformations of that tensor:
\begin{align}
\label{decomp}
{\rm tr}\,{\bm F}'
= \delta_{ij} F'_{ij}
= \delta_{ij} a_{ik} a_{jl} F_{kl}
= {\rm tr}\, {\bm F}\, .
\end{align}
By Schur's Lemma, irreducible tensors with respect to the orthogonal group are {\it traceless}
tensors having definite symmetry class (Young symmetry).
This decomposition is obtained as follows:
\begin{align}
\label{TL}
F_{i_1  \ldots i_n}^{[m]}
=
\cF^{[m]}_{i_1  \ldots i_n}
+ \sum\limits_{1 \leq r,s \leq n} \delta_{i_r i_s}\, 
F_{i_1 \ldots i_{r-1}i_{r+1}\ldots 
i_{s-1}i_{s+1}\ldots i_n}^{[m-2]}\,.
\end{align}
The tensors which appear under the sum in Eq.~(\ref{TL}) have rank $n-2$ and a
 Young pattern ${[m-2]}\subset {[m]}$ obtained by removing 
two boxes from the pattern. They may be decomposed again into traceless ones plus some
remainder, and so on. Therefore, a traceless tensor is obtained
from the original one by subtracting all the traces.

In general, there are two possible ways to construct $O(n)$-irreducible 
tensors. Either one symmetrizes the indices according to the
corresponding (standard) Young tableaux and afterwards subtracts 
all the traces, or one starts from tensors being already traceless 
and finally symmetrizes because this does not destroy the
tracelessness.
In this sense, the operations of Young symmetrization and subtraction of traces can be interchanged and are `commutative'.

\section{Traceless tensor of rank three}
\label{appendixA}
\setcounter{equation}{0}
\renewcommand{\theequation}{\thesection.\arabic{equation}}

Here, we decompose a general tensor of rank three $F_{ijk}$ into a {\it traceless tensor of rank three}
$\tl F_{ijk}$ and the {\it three traces}, given in terms of three vectors $H_k$, $K_j$ and $L_i$, 
according to (see, e.g., \citep{Hamer})
\begin{align}
\label{F-deco}
F_{ijk}=\tl F_{ijk}+\delta_{ij} H_k+
\delta_{ik}K_j+\delta_{jk}L_i\,.
\end{align}
We require that the tensor $\tl F_{ijk}$ be traceless and obtain from 
Eq.~(\ref{F-deco}):
\begin{align}
\label{A-gls}
\delta_{ij}
: \quad
F_{\cdot\cdot k}& =n H_k+K_k+L_k\nonumber\\
\delta_{ik}: \quad
F_{\cdot j\cdot}& = H_j+n K_j+L_j\\
\delta_{jk}: \quad
F_{i\cdot\cdot}& = H_i+K_i+n L_i\nonumber
\end{align}
where we use the notation
$F_{\cdot\cdot k}\equiv F_{iik}$, and $\delta_{ii}=n$.
Solving~(\ref{A-gls}), we find
\begin{align}
\label{A-sol}
H_i&=\frac{1}{(n+2)(n-1)}\big[(n+1)F_{\cdot\cdot i}-F_{\cdot i\cdot}
-F_{i\cdot\cdot}\big]\nonumber\\
K_i&=\frac{1}{(n+2)(n-1)}\left[-F_{\cdot\cdot i}+(n+1)F_{\cdot i\cdot}
-F_{i\cdot\cdot}\right]\\
L_i&=\frac{1}{(n+2)(n-1)}\left[-F_{\cdot\cdot i}-F_{\cdot i\cdot}
+(n+1)F_{i\cdot\cdot}\right]\nonumber\, .
\end{align}
Thus, a {\it traceless tensor of rank three} reads
\begin{align}
\label{F-tl}
\tl F_{ijk}=F_{ijk}-\delta_{ij} H_k-\delta_{ik}K_j-\delta_{jk}L_i
\end{align}
with Eq.~(\ref{A-sol}).
The decomposition (\ref{F-deco}) is unique \citep{Hamer,Weyl}.

\section{Irreducible pieces for gradient elasticity of form~I}
\label{appendixB}
\setcounter{equation}{0}
\renewcommand{\theequation}{\thesection.\arabic{equation}}

For completeness and convenience of the reader, we give the 
$O(n)$-irreducible pieces for gradient elasticity of form~I.

In {\it gradient elasticity of form~I}~\citep{Mindlin64}, 
the tensor of rank three, which is the {\it second gradient of the displacement
vector}, is used
\begin{align}
\label{eta-I}
\eta_{ijk}:=u_{k,ij}
\end{align}
and the two traces read
\begin{align}
\label{eta-tra1-I}
\eta_{\cdot\cdot k}=u_{k,ll}=\Delta u_k
\end{align}
and
\begin{align}
\label{eta-tra2-I}
\eta_{k\cdot\cdot}=u_{l,lk}\,.
\end{align}

Substituting Eqs.~(\ref{eta-I})--(\ref{eta-tra2-I})
into Eqs.~(\ref{T1-1}), (\ref{T2-1}), 
(\ref{T3-1}) and (\ref{T4-1}), 
the {\it four $O(n)$-irreducible pieces for gradient elasticity of form~I}
read
\begin{align}
\label{T1-1-I}
{}^{(1)}\eta_{ijk}&=
\frac{1}{3}\Big(u_{k,ij}+u_{i,jk}+u_{j,ki}\nonumber\\
&\hspace{15mm}
-\frac{1}{n+2}\Big(\delta_{ij}\big[2u_{l,lk}+u_{k,ll}\big]
+\delta_{ik}\big[2u_{l,lj }+u_{j,ll}\big]
+\delta_{jk}\big[2u_{l,li}+u_{i,ll}\big]\Big)
\Big)\\
%%\end{align}
%%\begin{align}
\label{T2-1-I}
{}^{(2)}\eta_{ijk}&=
\frac{1}{3}\Big(2 u_{k,ij}-u_{i,jk}-u_{j,ki}
\nonumber\\
&\hspace{15mm}
+\frac{1}{n-1}\Big(2\delta_{ij}\big[u_{l,lk}-u_{k,ll}\big]
-\delta_{ik}\big[u_{l,lj}-u_{j,ll}\big]
-\delta_{jk}\big[u_{l,li}-u_{i,ll}\big]\Big)
\Big)\\
%%\end{align}
%%\begin{align}
\label{T3-1-I}
{}^{(3)}\eta_{ijk}&=\frac{n}{(n+2)(n-1)}
\Big(\delta_{ik}\Big[u_{l,lj}-\frac{1}{n}\,u_{j,ll}\Big]
+\delta_{jk}\Big[u_{l,li}-\frac{1}{n}\,u_{i,ll}\Big]
-\frac{2}{n}\,\delta_{ij}\Big[u_{l,lk}-\frac{1}{n}\,u_{k,ll}\Big]\Big)
\Big)\\
%%\end{align}
%%\begin{align}
\label{T4-1-I}
{}^{(4)}\eta_{ijk}&=\frac{1}{n}\,\delta_{ij}\, u_{k,ll}\,.
\end{align}

\section{Irreducible decomposition of the tensor $F_{[ij]k}$}
\label{appendixD}
\setcounter{equation}{0}
\renewcommand{\theequation}{\thesection.\arabic{equation}}

We now give the decomposition of the tensor
$\kappa_{ijk}:=F_{[ij]k}$ ($\kappa_{ijk}=-\kappa_{jik}$) into its
$O(n)$-irreducible pieces.

For the tensor $\kappa_{ijk}$ there are only two non-vanishing Young tableaux if
$n>2$, namely 
\begin{eqnarray}\label{tab4}
 \Yvcentermath1\young({i},{j})\otimes\young(k)
 &=&\Yvcentermath1 \beta\,\young({i}{k},{j})
 \Yvcentermath1 + \gamma\,\young({i},{j},{k}) 
%\end{align}
\end{eqnarray}
\begin{align}
\label{deco-YT-k}
\qquad\kappa_{[ij]k}&=\frac{4}{3}\, \kappa_{[j(i]k)}+
\kappa_{[ijk]}\,
\end{align}
%%where $\alpha$ is a normalization factor.
%%%% $\alpha=1/12$.
with the two tensor pieces
\begin{align}
\label{kappa-P-gl}
{}_{{\text{P}}}\kappa_{ijk}:=
\frac{4}{3}\, \kappa_{[j(i]k)}&=
\frac{1}{3}\big(2 \kappa_{ijk}-\kappa_{jki}-\kappa_{kij}\big)\\
\label{kappa-A-gl}
{}_{\text{A}}\kappa_{ijk}:=
\kappa_{[ijk]}&=\frac{1}{3}\big(\kappa_{ijk}+\kappa_{jki}+\kappa_{kij}\big)
\end{align}
which are irreducible with respect to $GL(n)$.
Moreover, ${}_{\text{A}}\kappa_{ijk}$ is already irreducible 
and ${}_{{\text{P}}}\kappa_{ijk}$ is reducible with respect to $O(n)$.
Note that \citet{Toupin62} called ${}_{{\text{P}}}\kappa_{ijk}$ the principal part.

Finally, 
the tensor $\kappa_{ijk}$ can be decomposed into 
{\it three irreducible pieces} under the orthogonal group $O(n)$
if we split the  principal part ${}_{{\text{P}}}\kappa_{ijk}$ into its traceless and trace parts.
For $n>2$, we may write its three irreducible pieces
according to 
\begin{align}
&\kappa_{ijk}={}^{(1)}\kappa_{ijk}+{}^{(2)}\kappa_{ijk}+{}^{(3)}\kappa_{ijk}
\end{align}
with the number of independent tensor components
\begin{align}
&\frac{n^2(n-1)}{2}=\frac{n(n^2-4)}{3}+n+\frac{n(n-1)(n-2)}{6}\,
\end{align}
and the three $O(n)$-irreducible pieces are given by
\begin{align}
\label{kappa-1}
{}^{(1)}\kappa_{ijk}&:=
{}_{\text{P}}\tl\kappa_{ijk}\equiv\frac{4}{3}\, \tl\kappa_{[j(i]k)}
%\nonumber\\
=\frac{1}{3}\big(2 \kappa_{ijk}-\kappa_{jki}-\kappa_{kij}\big)
+\frac{1}{n-1}\,\big(\delta_{ik}\kappa_{j\cdot\cdot}-\delta_{jk}\kappa_{i\cdot\cdot}\big)\\
\label{kappa-2}
{}^{(2)}\kappa_{ijk}&=\frac{1}{n-1}\,\big(\delta_{jk}\kappa_{i\cdot\cdot}-\delta_{ik}\kappa_{j\cdot\cdot}\big)\\
{}^{(3)}\kappa_{ijk}&:={}_{{\text{A}}}\kappa_{ijk}
%%%=\kappa_{[ijk]}
%\nonumber\\
=\frac{1}{3}\big(\kappa_{ijk}+\kappa_{jki}+\kappa_{kij}\big)\,.
\label{kappa-3}
\end{align}
The irreducible pieces are canonical. 
%%They satisfy
%%\begin{align}
%%&{}^{(1)}\kappa_{i\cdot\cdot }={}^{(3)}\kappa_{i\cdot\cdot}=0
%%\end{align}
%%and 
%%\begin{align}
%%{}^{(2)}\kappa_{i\cdot\cdot}=\kappa_{i\cdot\cdot }\,.
%\end{align}
Under the group $SO(3)$, the pieces ${}^{(1)}\kappa_{ijk}$, ${}^{(2)}\kappa_{ijk}$, ${}^{(3)}\kappa_{ijk}$ are spin-2, spin-1, spin-0 fields, respectively. 
For $n=2$, the tensor $\kappa_{ijk}$ is already irreducible.

For $n=3$, we may define the {\it dual tensor} of $\kappa_{ijk}$ according to
\begin{align}
\kappa_{ij}:=\frac{1}{2}\,\epsilon_{klj} \kappa_{kli}
\end{align}
and the inverse relation reads
\begin{align}
\kappa_{jki}=\epsilon_{jkn}\kappa_{in}\,.
\end{align}
For the irreducible pieces, it holds for every irreducible piece:
${}^{(I)}\kappa_{ij}=\frac{1}{2}\epsilon_{klj} {}^{(I)}\kappa_{kli}$
for $I=1,2,3$.
Then the three dual $O(3)$-irreducible pieces are 
\begin{align}
&\kappa_{ij}={}^{(1)}\kappa_{ij}+{}^{(2)}\kappa_{ij}+{}^{(3)}\kappa_{ij}
\end{align}
where
\begin{align}
\label{kappa-1-2}
{}^{(1)}\kappa_{ij}&=\kappa_{(ij)}-\frac{1}{3}\, \delta_{ij}\kappa_{ll}
={\kappa\quer}_{\!\! (ij)}\\
\label{kappa-2-2}
{}^{(2)}\kappa_{ij}&=\kappa_{[ij]}\\
{}^{(3)}\kappa_{ij}&=\frac{1}{3}\, \delta_{ij}\kappa_{ll}\,.
\label{kappa-3-2}
\end{align}
Thus, ${}^{(1)}\kappa_{ij}$ is the symmetric and traceless part 
(or symmetric deviator) of $\kappa_{ij}$,
 ${}^{(2)}\kappa_{ij}$ is the antisymmetric part of $\kappa_{ij}$,
and ${}^{(3)}\kappa_{ij}$ is the trace part (or spherical part) of $\kappa_{ij}$. 
The pieces ${}^{(1)}\kappa_{ij}$, ${}^{(2)}\kappa_{ij}$, ${}^{(3)}\kappa_{ij}$ are spin-2, spin-1, spin-0 fields, respectively.

\section{Irreducible decomposition of a general tensor of rank three $F_{ijk}$}
\label{appendixE}
\setcounter{equation}{0}
\renewcommand{\theequation}{\thesection.\arabic{equation}}

As an important side-result of this paper, 
we give the {\it irreducible decomposition 
of a general tensor of rank three} $F_{ijk}$ with respect to $O(n)$. 
The first step is to decompose  the general tensor $F_{ijk}$  in the indices $i$ and $j$ into symmetric and antisymmetric parts.
In the second step, we use the irreducible decomposition of 
these two tensors. 
For $n>2$, a general tensor of rank three, $F_{ijk}$, can be uniquely
decomposed into {\it seven $O(n)$-irreducible pieces} according to 
\begin{align}
\label{F-deco-0}
F_{ijk}&=F_{(ij)k}+F_{[ij]k}
%%\nonumber\\&
=\eta_{ijk}+\kappa_{ijk}\nonumber\\
&={}^{(1)}\eta_{ijk}+{}^{(2)}\eta_{ijk}+{}^{(3)}\eta_{ijk}+{}^{(4)}\eta_{ijk}+{}^{(1)}\kappa_{ijk}+{}^{(2)}\kappa_{ijk}+{}^{(3)}\kappa_{ijk}
\end{align}
with the seven irreducible tensor pieces given by
Eqs.~(\ref{T1-2}), (\ref{T2-2}), (\ref{T3-2}), (\ref{T4-1}), (\ref{kappa-1}), (\ref{kappa-2}), (\ref{kappa-3}). 
For instance, the irreducible decomposition (\ref{F-deco-0}) can be used in micromorphic elasticity for the micro-deformation gradient tensor (see, e.g., \citep{Mindlin64,Eringen}).

\end{appendix}


\begin{thebibliography}{}

\bibitem[Auffray(2013)]{Auffray13}
        N.~Auffray,
	On the algebraic structure of isotropic generalized elasticity theories,
	Mathematics and Mechanics of Solids~{\bf 20}, 565--581 (2015).

\bibitem[Auffray(2015)]{Auffray15}
         N.~Auffray,
	 On the isotropic moduli of 2D strain-gradient elasticity,
	 Continuum Mechanics and Thermodynamics~{\bf 27}, 5--19 (2015).

\bibitem[Barut and Raczka(1977)]{BR}
         A.O.~Barut and R.~Raczka,
	 {Theory of Group Representations and Applications},
	 PWN -- Polish Scientific Publishers, Warszawa (1977).

\bibitem[Bertram(2005)]{Bertram}
        	A.~Bertram, 
	{Elasticity and Plasticity of Large Deformations}, 
	Springer, Berlin (2005).

\bibitem[Boerner(1955)]{Boerner}
        H.~Boerner, 
	{Darstellungen von Gruppen}, Springer-Verlag, Berlin (1967);
	Representations of Groups, North-Holland, Amsterdam (1970).

\bibitem[Chadwick(1976)]{Chadwick}
	P.~Chadwick,
        Continuum Mechanics: Concise Theory and Problems,
        George Allen and Unwin, London (1976).

\bibitem[Eringen(1999)]{Eringen}
        	A.C.~Eringen, 
	{Microcontinuum Field Theories I: Foundations and Solids}, 
	Springer, New York (1999).

\bibitem[Fleck and  Hutchinson(1997)]{Fleck}
        N.A.~Fleck and J.W.~ Hutchinson,
	Strain gradient plasticity,
	Advances in Applied Mechanics~{\bf 33}, 296--361 (1997).

\bibitem[Gao et al.(1999)]{Gao}
        H.~Gao, Y.~Huang, W.D.~Nix, and J.W.~Hutchinson,
        Mechanism-based strain gradient plasticity -- I. Theory,
        Journal of the Mechanics and Physics of Solids~{\bf 47}, 
        1239--1263 (1999).

\bibitem[Geyer et al.(1999))]{GLR99}
         B.~Geyer, M.~Lazar, and D.~Robaschik, 
	 Decomposition of non-local light-cone operators into harmonic operators of definite twist,
	 Nuclear Physics~B~{\bf 559}, 339--377 (1999).

\bibitem[Geyer and Lazar(2000)]{GL00}
         B.~Geyer and M.~Lazar, 
	 Twist decomposition of nonlocal light-cone operators II: general tensors of 2nd rank,
	 Nuclear Physics B~{\bf 581}, 341--390 (2000).

\bibitem[Gl\"uge et al.(2015)]{GK}
        R.~Gl\"uge, J. Kalisch, and A.~Bertram,
	The eigenmodes in isotropic strain gradient elasticity,
	in: 
	Generalized Continua as Models for Classical and Advanced Materials, 
        pp.~163--178,
	Eds.: H. Altenbach and S. Forest, Springer (2016). 

\bibitem[Hamermesh(1962)]{Hamer}
        M.~Hamermesh,
	{Group Theory and Its Application to Physical Problems},
	Addison-Wesley, London (1962);
	Dover, New York (1989).

\bibitem[Hehl et al.(1995)]{Hehl95} 
        F.W.~Hehl, J.D.~McCrea, E.W.~Mielke, and Y.~Ne'eman, 
        {Metric--affine gauge theory of gravity: Field equations, Noether
        identities, world spinors, and breaking of dilation invariance},
         Physics Reports {\bf 258}, 1--171 (1995).

\bibitem[Itin and Hehl(2015)]{IH} 
        Y.~Itin and F.W.~Hehl, 
	Irreducible decompositions of the elasticity tensor under the linear and orthogonal groups and their physical consequences,
	Journal of Physics: Conference Series~{\bf 597}, 012046 (2015). 
 
\bibitem[Landau and Lifshitz(1970)]{LL}
          L.D.~Landau and  E.M.~Lifschitz, 
         Theory of Elasticity: Volume 7 (Course of Theoretical Physics), 
          Pergamon Press Ltd., Oxford (1970).
 
\bibitem[Lazar(2002)]{Lazar02}
         M.~Lazar,
	An elastoplastic theory of dislocations as a physical field theory with torsion,
         Journal of Physics A: Mathematical and General~{\bf 35}, 1983--2004 (2002).
 
 \bibitem[Lazar and Maugin(2005)]{LM05} 
        M.~Lazar and G.A.~Maugin, 
        Nonsingular stress and strain fields of dislocations and disclinations
        in first strain gradient elasticity,
        %%Int. J. Engng. Sci.
       International Journal of Engineering Science~{\bf 43}, 1157--1184 (2005).

\bibitem[Lazar et al.(2005)]{LMA05} 
        M.~Lazar, G.A.~Maugin, and E.C.~Aifantis,
        On dislocations in a special class of generalized elasticity,
        physica status solidi~(b)~{\bf 242}, 2365--2390 (2005).


\bibitem[Lazar and Anastassidis(2009)]{LA09} 
	M.~Lazar and C.~Anastassiadis,
	Gauge theory of dislocations: static solutions of screw and edge dislocations,
	Philosophical Magazine~{\bf 89}, 199--231 (2009).

\bibitem[Lazar(2013)]{Lazar13}
        M.~Lazar, 
        The fundamentals of non-singular dislocations in the theory of
        gradient elasticity: Dislocation loops and straight dislocations,
        %%Int. J. Solids Struct.
        International Journal of Solids and Structures~{\bf 50}, 352--362 (2013).

\bibitem[Lazar(2014)]{Lazar14}
        M.~Lazar, 
        On gradient field theories:
        gradient magnetostatics and gradient elasticity, 
        %%Phil. Mag.
        Philosophical Magazine~{\bf 94}, 2840--2874 (2014).

\bibitem[Mindlin(1964)]{Mindlin64}
        R.D.~Mindlin, 
        Micro-structure in linear elasticity,
        Archive for Rational Mechanics and Analysis~{\bf 16},     
        %%Arch. Rational. Mech. Anal.~{\bf 16}, 
        51--78 (1964).

\bibitem[Mindlin and Eshel(1968)]{Mindlin68}
         R.D.~Mindlin and N.N.~Eshel, 
        On first strain-gradient theories in linear elasticity,
        %%Int. J. Solids Struct.
        International Journal of Solids and Structures~{\bf 4}, 109--124 (1968).

\bibitem[Monchiet and Bonnet(2011)]{MB}
         V.~Monchiet and G.~Bonnet,
         Inversion of higher order isotropic tensors with minor symmetries and solution of higher order heterogeneity problems,
         Proceedings of the Royal Society A~{\bf 467},  314--332 (2011).

\bibitem[Neff et al.(2015)]{Neff} 
	P.~Neff, I.-D.~Ghiba,  M.~Lazar, and A.~Madeo,
	The relaxed linear  micromorphic continuum: well-posedness of the static problem and relations to the gauge theory of dislocations,  
	The Quarterly Journal of Mechanics and Applied Mathematics~{\bf 68}, 53--84 (2015).

\bibitem[Schouten(1951)]{Schouten} 
        J.A.~Schouten, 
        Tensor Analysis for Physicits,
        Oxford University Press, Oxford (1951).

\bibitem[Sexl and Urbantke(2001)]{Sexl}
        R.U.~Sexl and H.K.~Urbantke,
        Relativity, Groups, Particles: Special Relativity and Relativistic
        Symmetry in Field and Particle Physics,
        Springer, Wien (2001). 

\bibitem[Shodja et al.(2013)]{Shodja}
        H.M.~Shodja, A.~Zaheri, and A.~Tehranchi, 
        Ab initio calculations of characteristic lengths of crystalline 
        materials in first strain gradient elasticity,
        %%Mech. Mater.
        Mechanics of Materials~{\bf 61}, 73--78 (2013).

\bibitem[Toupin(1962)]{Toupin62}
        R.A.~Toupin,
        Elastic materials with couple-stresses, 
        Archive for Rational Mechanics and Analysis~{\bf 11},     
        %%Arch. Rational. Mech. Anal.~{\bf 16}, 
        385--414 (1962).


\bibitem[Wade and Bruck(1944)]{Wade44}
        T.L.~Wade and R.H.~Bruck,
        Types of Symmetries, 
        The American Mathematical Monthly~{\bf 51}, 123--129 (1944). 

\bibitem[Weyl(1931)]{Weyl1} 
        H.~Weyl, 
        Gruppentheorie und Quantenmechanik,
        S. Hirzel,  Leipig (1931); 
        The Group Theory and Quantum Mechanics,
        Dover, New York (1955).

\bibitem[Weyl(1952)]{Weyl} 
        H.~Weyl, 
        The Classical Groups: Their Invariants and Representations,
        Princeton University Press Princeton (1946).

\end{thebibliography}
\end{document}